\documentclass{aa}

\usepackage{epsf,graphics}

\begin{document}
  \thesaurus{12      
          (09.04.1;  
           09.07.1)} 

\title{Interstellar nanodiamonds:
       the carriers of mid-infrared emission bands?}

\titlerunning{Interstellar nanodiamonds}

\author{A.P.\,Jones \thanks{\emph{Email address:} ant@ias.fr}
 and L.\,d'Hendecourt}

\institute{Institut d'Astrophysique Spatiale (CNRS), Universit\'e
           Paris XI - B\^atiment 121, 91405 Orsay Cedex, France}

\date{Received / Accepted  : }
         
\maketitle


\begin{abstract}
In this paper we pursue the natural consequences of the structure of
nanodiamonds and their surface relaxation and reconstruction to
surfaces exhibiting sp$^2$ carbon atoms. We show that in the
interstellar radiation field nanodiamonds can be stochastically heated
to temperatures as high as 1000~K and give rise to discrete emission
bands associated with the surface structures. We therefore speculate
that nanodiamonds can make a significant contribution towards the
$3-15~\mu$m unidentified mid-infrared emission bands.
\end{abstract}

\section{Introduction}

Diamonds were first proposed as a component of interstellar dust by
Saslaw \& Gaustad in 1969. However, it was only in 1987 that presolar
nanometre-sized diamond grains (nanodiamonds) were first detected, not
in the interstellar medium, but in the extracted fractions from
primitive meteorites (Lewis et al. 1987). These diamond grains are
composed of almost pure carbon with less than 1 \% by mass of
impurities with atomic masses greater than 6 (Bernatowicz et al.
1989). They contain isotopic anomalies, e.g., $^{15}$N depletions and
likely $p-$ and $r-$process Xe compositions, and enhanced D/H ratios,
that clearly indicate a presolar origin (e.g., Anders \& Zinner 1993).

Nanodiamonds are, by over two orders of magni\-tude in mass, the most
abundant presolar grains recovered from primitive meteorites.  Their
euhedral morphologies (well-formed crystal shapes) indicate that they
have undergone little alteration (Daulton et al. 1996).

It has been postulated that interstellar diamonds will be observable
only with great difficulty (Lewis et al. 1989). This is, however, a
conclusion inferred principally from the extrapolation of the bulk
diamond properties to nanometre size-scales.  However, as Bernatowicz
et al. (1989) caution, attempting to model interstellar diamonds as
smaller versions of macroscopic diamond will lead to errors.

The search for observable intrinsic nanodiamond properties in the
laboratory has been hampered by their high surface to volume ratio and
active surface chemistry (Hill et al. 1998). The infrared spectra of
the extracted nanodiamonds are dominated by surface adsorbates
introduced during laboratory extraction and, as Hill et al. (1998)
pointed out, this is clearly the case with many published presolar
nanodiamond spectra (e.g., Koike et al. 1995; Mutschke et al. 1995;
Andersen et al. 1998). This active surface chemistry, while aiding
their extraction (e.g., Anders \& Zinner 1993), has clearly been a
hindrance to the search for observable nanodiamond signatures.

Allamandola et al. (1992,1993) suggested that an absorption band at
3.47~$\mu$m could be due to a tertiary carbon CH stretching mode on
the surfaces of nanodiamonds. However, this band is probably not
unique to nanodiamonds as it seems that it could have other origins
(Brooke et al. 1996). Recently, Guillois et al. (1999) showed that
infrared emission features observed at 3.43 and 3.53~$\mu$m in two
sources are well-matched by surface CH groups on $\sim 50$~nm radius
diamond particles.

Lewis et al. (1989) suggested that diamond grains may be detectable in
emission, and Hill et al. (1998) noted that they may only be
observable in emission. Hill et al. (1998) showed that a diamond
feature at $\sim 21~\mu$m was consistent with the unidentified
emission feature seen at this wavelength in some carbon-rich
proto\-planetary nebulae (PPNe).

The measurement of the intrinsic infrared to ultraviolet spectra of
nanometre-sized diamond particles and their physical properties is
central to determining the observability of interstellar
nanodiamonds. The extrapolation of bulk diamond properties is clearly
not appropriate because a large fraction of the atoms are in or near
the surface where the structure is different from that of the bulk
material. In this paper we model the structural properties of
nanometre-sized diamonds and discuss their infrared properties.

\section{The formation of nanodiamonds}

Presolar nanodiamonds were originally isolated on the basis of their
anomalous Xenon isotopic ratios, the so-called Xe-HL meteoritic
component (Lewis et al. 1987). This isotopic component seems to
indicate a supernova origin for the nanodiamonds. However, their
formation in supernovae is seemingly inconsistent with the requirement
for a carbon-rich environment (e.g., Lewis et al. 1989).

Daulton et al. (1996) showed that the twinning structures and the
pau\-city of dislocations within the bulk structure of meteoritic
nanodiamonds imply rapid, isotropic growth and vapour phase processes
for their formation. The measured log-normal nanodiamond size
distribution also seems to be consistent with a growth mechanism
rather than a fragmentation formation process, and indicates minimal
contribution from sputtering and erosion processes (Lewis et al. 1989;
Daulton et al. 1996).

In the chemical vapour deposition (CVD) formation of diamond in the
laboratory, growth occurs at the gas-solid interface and both atomic
hydrogen and oxygen play important roles in the preferential erosion
of non-diamond carbon, principally graphitic carbon, from the growth
surface (e.g., Angus et al. 1989).  The synthesis of diamond in the
laboratory requires $0.8 \leq {\rm C/O} \leq 1.3$ (Bachmann et
al. 1991) which is entirely consistent with the observed C/O ratios
for carbon stars (Lambert et al. 1986).

Although diamond growth by CVD processes is now reasonably well
understood (e.g., Angus et al. 1989), the initial nucleation
process is not clear. It is, however, clear that the graphite/diamond
interface plays a very fundamental role in CVD diamond formation and
growth (e.g., Lambrecht et al. 1993). Lambrecht et al. (1993) have
shown that the initial formation steps in diamond growth can occur by
nucleation on a graphite surface through a hydrogenation process.
Interestingly, C$_{60}$ and C$_{70}$ fullerenes have also been shown
to act as diamond nucleation centres under CVD conditions and to
induce high diamond nucleation rates (Meilunas et al. 1991). These
very fundamental processes therefore point towards a very intimate
connection between sp$^3$ diamond and sp$^2$ graphite surfaces and
interfaces.

Nanodiamonds can also be formed by the intense ultra\-violet
irradiation of graphitic and amor\-phous carbon particles (Fedoseev et
al. 1983; Alam et al. 1989). Such processes in the astrophysical
context have been invoked to explain diamond formation in supernovae
(Nuth \& Allen 1992; Ozima \& Mochizuki 1993). However, the thermal
processing of nanodiamonds at temperatures of the order of $1400 -
1800$~K leads to the loss of the diamond structure and their
conversion to onion-like graphitic particles (Kuznetsov et al. 1994).
It has been noted that, at relatively low temperature, pressure
exerted on graphite along the direction of the crystallographic
$c$-axis can induce the formation of lonsdaleite (see discussion of
diamond polytypes by Daulton et al. 1996). The presence of lonsdaleite
in meteorites may therefore indicate the shock transformation of
graphite in the interstellar medium (e.g., Tielens et
al. 1987). However, this diamond polytype can also have a low-pressure
origin (Frenklach et al. 1989) and so the evidence for shock-produced
interstellar nanodiamonds is equivocal. The contribution of
shock-produced nanodiamonds to the meteoritic nanodiamond budget
seems, in any event, to be small (Daulton et al. 1996).

\section{Presolar nanodiamonds: structure, size and shape}

The cubic form of diamond has a structure consisting of two
interleaved face-centred cubic (fcc) lattices, but diamond also has a
hexagonal structural form (lonsdaleite) which is thought to be the
less stable of the two. The presolar nanodiamonds clearly show the
characteristics expected of the perfect cubic diamond fcc lattice
(e.g., Bernatowicz et al. 1990; Daulton et al. 1996). However, it has
long been known that perfect diamond (111) surfaces can reconstruct to
(2$\times$1) and/or (2$\times$2) sp$^2$-dominated structures (e.g.,
Vanderbilt \& Louie 1984; Mitsuda et al. 1991).

In several studies (Blake et al. 1988; Bernatowicz et al. 1990;
Dorschner et al. 1996; Hill 1998) an observed $1s
\rightarrow \pi^\ast$ transition in the electron energy loss spectra
(EELS) of meteoritic nanodiamonds was taken to be evidence for the
presence of a $\pi$-bonded carbon component on the surfaces. The
observation of this intimately related sp$^2$ component in these
studies was taken to be strong evidence for surface reconstruction in
nanometre-sized diamonds and was assumed to be a natural consequence
of the small particle sizes. The corresponding sp$^2$/sp$^3$ ratios
were found to be $\sim 25$\% (Blake et al. 1988), $> 25$\%
(Bernatowicz et al. 1990), $\sim 20-27$\% (Dorschner et al. 1996) and
$\sim 10-20$\% (Hill 1998).

Bernatowicz et al. (1990) found that the surface material most
consistent with their EELS data was a form of hydrogenated amorphous
carbon (a-C:H), rather than graphite or amorphous carbon (a-C).  In
partic\-ular, they not\-ed the constancy of the proportion of
$\pi$-bonded carbon (sp$^2$) to fcc carbon (sp$^3$) in the meteoritic
nanodiamonds.  They modelled their EELS spectra using diamond mantled
with graphite, a-C and a-C:H, and found a best fit for 46 \% by volume
of fcc diamond and 54 \% by volume of a-C:H (with $35-60$ \% of
hydrogen) which was assumed to be in the form of a coating on the
presolar diamonds. These data are consistent with the fact that the
nanodiamond residues are brown and transparent, i.e., neither black,
as expected for graphite and a-C, nor completely transparent, as would
be the case for pure uncoated diamond. In addition, they found that
the specific density of the nanodiamonds (2.22--2.33~g~cm$^{-3}$,
Lewis et al. 1987) was entirely consistent with this two component
diamond core and a-C:H coating.

Daulton et al. (1996) have shown that the shape of the presolar
nanodiamonds is variable but that well-faceted particles (euhedral
morphologies) predominate. For the examin\-ed meteoritic
nano\-diamonds they find significant depart\-ures from spherical
geometry, log-normal size distributions and a median radius of
$\sim~1.4$~nm.

\section{A model for the structural properties of nanometre-sized diamonds}

The fraction of surface and near surface carbon atoms in a
nanometre-sized diamond particle will be only weakly dependent upon
the particle size. This is simply because the number of surface atoms
varies as the square of the particle radius while the total number of
atoms varies as the cube of the radius. The ratio of the number of
surface atoms to the total number of atoms is then simply proportional
to the inverse of the particle radius. Thus, a factor of eight
increase in the nanodiamond mass (or a doubling of the radius) will
lead to only a factor of two decrease in the fraction of surface
atoms. For the $\sim 1-2$~nm radius presolar diamonds the surface atom
fraction thus only varies by a small factor. For the measured
meteoritic nanodiamonds the size distribution seems to be relatively
independent of the source meteorite (e.g., Bernatowicz et al. 1990;
Daulton et al. 1996) and so the surface atom fraction for presolar
diamonds is essentially constant.

\begin{figure}
 \epsfxsize=8.5cm
 \epsfysize=7.5cm
 \epsffile{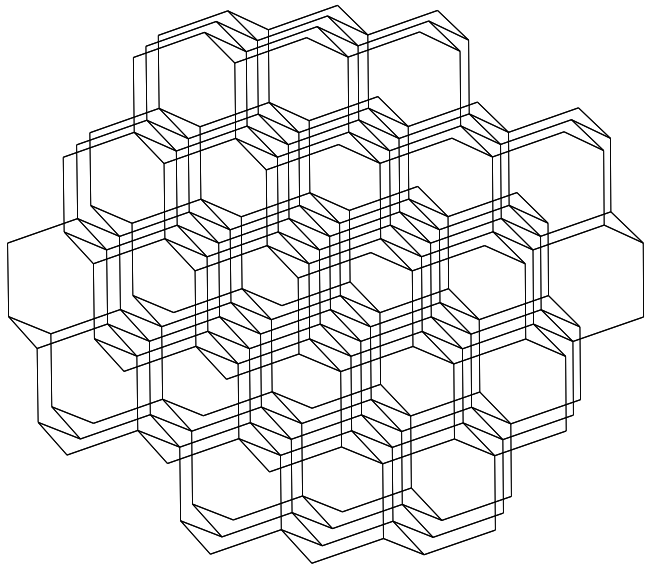}
 \caption{A 226 carbon atom diamond nanoparticle. In this stick figure
 of the three-dimensional particle the end-points of each line, which
 represents a C--C bond, mark the position of a carbon atom in the fcc
 diamond lattice. The upper, lower, upper right and lower left faces
 are predominantly diamond (111)-type crystal planes. The upper left
 and lower right are predominantly diamond (100)-type crystal
 planes. The front and back faces are composed of (111)-type and
 (100)-type crystal planes (as per the left and right faces seen
 edge-on).}
 \label{small}
\end{figure}

We have computed the structural properties of approximately spherical
nanodiamond particles using a cubic grid filled with sp$^3$ carbon
atoms according to the cubic fcc diamond structure. We then generated
a nanodiamond particle by selecting only those carbon atoms within a
defined radius. In order to model compact structures we remove from
the particle any carbon atoms that are only attached to one other
carbon atom (the primary, 1$^\circ$, carbon atom sites). As a result
the particles exhibit spheroidal shapes showing some reasonably
well-defined diamond (111)-type and (100)-type crystallographic
faces. In Fig.~\ref{small} we show an example of a 226 carbon atom
diamond particle of radius $\approx 0.6$~nm. The illustrated particle
exhibits (111)-type and (100)-type crystallographic faces which
contain approximately equal numbers of carbon atoms. Although, as we
show below the (111)-type faces are more abundant in these particles.

In Table~1 we show the results for nanodiamond structures with radii
in the $\approx 0.6-2.1$~nm range which covers the typical presolar
nanodiamond sizes ($\sim 1-2$~nm). The data in Table~1 show the number
of secondary, 2$^\circ$, tertiary, 3$^\circ$, and quaternary,
4$^\circ$, (bonded to 2, 3 and 4 other atoms, respectively), sp$^3$
carbon atoms as a function of the particle size. Within these
structures only atoms bonded to four others (4$^\circ$ atom sites) can
be within the bulk, and those bonded to less than four (2$^\circ$ and
3$^\circ$ atom sites) must be surface atoms in the absence of internal
vacancies. The surface to total carbon atom ratio is then simply the
number of 2$^\circ$ and 3$^\circ$ atoms divided by the total number of
atoms. We also note that the (111)-type faces contain only 3$^\circ$
carbon atoms and that the (100)-type faces contain only 2$^\circ$
carbon atoms (Fig.~\ref{small}). From the total number of 2$^\circ$
and 3$^\circ$ carbon atoms in the particles (Table~1) we therefore
conclude that in these particles there are approximately twice as many
atoms in the (111)-type crystallographic face structures as in the
(100)-type faces. This preference may be further accentuated in the
meteoritic diamonds which show euhedral morphologies and significant
departures from spherical shapes.

\begin{table}
\caption{Model results for typical meteoritic nanodiamond-sized
 structures. The right hand column gives the ratio of the number of
 surface atoms to the total number of carbon atoms in the particle.}
\begin{tabular}{|c|c|c|c|c|c|}
\hline 
 \multicolumn{1}{|c}{$\approx$ Radius}               &
 \multicolumn{4}{|c|}{Number of sp$^3$ carbon atoms} &
 \multicolumn{1}{|c|}{Fraction}                      \\
 \multicolumn{1}{|c}{( nm )} &
 \multicolumn{1}{|c|}{Total}           &
 \multicolumn{1}{|c|}{2$^\circ$}       &
 \multicolumn{1}{|c|}{3$^\circ$}       &
 \multicolumn{1}{|c|}{4$^\circ$}       &
 \multicolumn{1}{|c|}{in surface}      \\
\hline 
\hline
  0.6  &   226  &   42  &   60  &   124  &  0.45  \\
  1.0  &   678  &   78  &  140  &   460  &  0.32  \\
  1.2  &  1274  &  120  &  218  &   936  &  0.27  \\
  1.4  &  2124  &  180  &  304  &  1640  &  0.23  \\
  1.7  &  3275  &  225  &  436  &  2614  &  0.20  \\
  1.9  &  4851  &  270  &  584  &  3997  &  0.18  \\
  2.1  &  6728  &  336  &  740  &  5652  &  0.16  \\
\hline 
\end{tabular}
\end{table}

Obviously the dangling bonds on the surfaces of these small diamond
particles (due to carbon atoms bonded to less than four others) can be
terminated with hydrogen atoms. If this is the case then on the grain
surface we will find only sp$^3$ CH$_2$ (2$^\circ$) and CH (3$^\circ$)
groups. Recall that we have deliberately excluded primary, 1$^\circ$,
carbon atoms and hence there can be no CH$_3$ groups on the generated
particles. The surface hydrocarbon groups are generally isolated from
one another by at least one carbon atom of higher order on the regular
crystal faces, i.e., 2$^\circ$ CH$_2$ groups are separated by
3$^\circ$ and 4$^\circ$ sites, and 3$^\circ$ CH groups are separated
by 4$^\circ$ sites. However, along the particle edges this is no
longer true because many adjacent carbon atoms are hydrogenated.

\begin{figure}
 \epsfxsize=8.5cm
 \epsfysize=8.5cm
 \epsffile{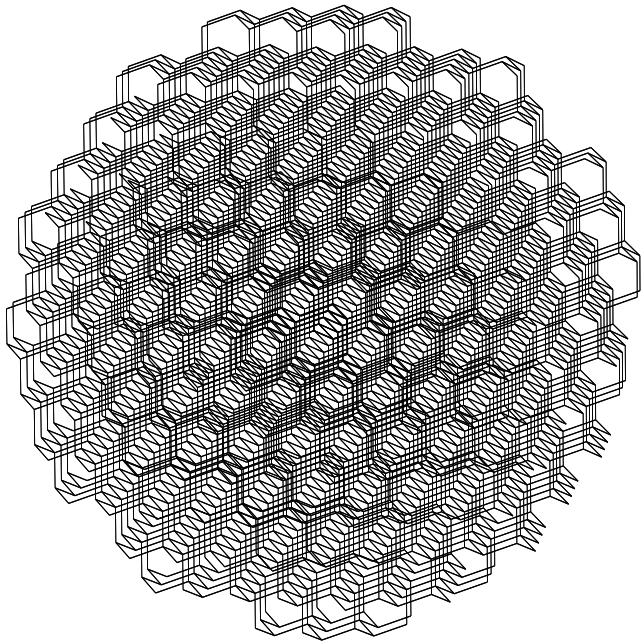}
 \caption{An idealised spheroidal meteoritic diamond nanoparticle
 with perfect fcc structure and radius $\approx 1.4$~nm. Again, in this
 stick figure of the three-dimensional particle the end-points of each
 line, which represents a C--C bond, mark the position of a carbon atom
 in the fcc diamond lattice. In the profile of this particle it can be
 seen that the upper, lower, upper right and lower left faces are
 dominated diamond (111)-type crystal planes, and the upper left and
 lower right are diamond (100)-type crystal planes.}
 \label{big}
\end{figure}

The formulae for the diamond particles in Table~1 are then;
C$_{226}$H$_{144}$, C$_{678}$H$_{296}$, C$_{1274}$H$_{458}$,
C$_{2124}$H$_{664}$, C$_{3275}$H$_{886}$, C$_{4851}$H$_{1124}$ and
C$_{6728}$H$_{1412}$. The H/C ratios lie between 0.64 for the smallest
particles and 0.21 for the largest. In Fig.~\ref{big} we show the
structure of the perfect fcc nanodiamond particle of radius $\approx
1.4$~nm (Table~1). This particle corresponds to approximately the mean
meteoritic diamond dimensions (Daulton et al. 1996).

\begin{figure}
 \epsfxsize=8.75cm
 \epsfysize=8.0cm
 \epsffile{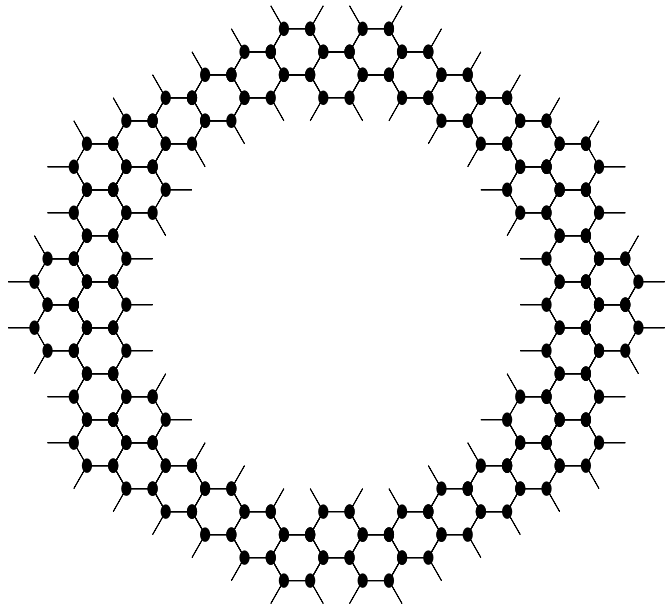}
 \caption{Schematic cross section of the reconstructed sp$^2$ surface
 layer of a $\sim 1.5$~nm radius nanodiamond particle with an
 ellipticity of 0.95 typical of the meteoritic nanodiamonds
 (Daulton et al. 1996). This structure is analogous to the periphery
 of a large compact PAH structure. The filled circles represent
 surface and near-surface sp$^2$ carbon atoms, and the lines between
 them represent C--C bonds. The lines pointing inward represent
 connections to the core diamond structure. Each line pointing outward
 represents a surface CH bond. In this schematic section through the
 nanoparticle it can be seen that there are only solo and duo CH
 groups present in about equal numbers.}
 \label{section}
\end{figure}

The discussion, so far, has considered only the case of the perfect
fcc diamond lattice terminated with sp$^3$ CH bonds. However, the
evidence presented in \S~3 clearly shows that diamond surfaces can
reconstruct to structures that are dominated by sp$^2$ carbon
atoms. It is thus necessary to consider the consequences for this
surface reconstruction in the above structural model. In the absence
of overlying layers of a different carbon composition, which we
consider below, a limit to the effects of surface reconstruction can
be estimated by assuming that all surface 2$^\circ$ and 3$^\circ$ atom
sites become sp$^2$ carbon atom sites. In this hypothetical sp$^2$
limit we can assume that all surface CC bonds are entirely olefinic or
aromatic and identical in character, and that this results from the
breaking of the bonds between 3$^\circ$ sites only. In
Fig.~\ref{section} we show a schematic cross section of the
reconstructed surface of a nanodiamond. As Figs.~\ref{small},
\ref{big} and \ref{section} show, the reconstruction of the surface to
sp$^2$ will result in the majority of the surface olefinic or aromatic
CH bonds having solo and duo CH configurations with perhaps some trio
CH groups present at steps and edges on the surface. The fraction of
the duo and trio sites may diminish somewhat as the particle size
increases, but the solo/(duo+trio) CH ratio will not be particularly
sensitive to the particle size. It is intriguing to note that about
half of the surface CH bonds could be solo (see Fig.~\ref{section})
and that this is relatively independent of the particle size.

Bernatowicz et al. (1990) find that a surface layer of a-C:H with
about an equal volume to the diamond grain centre is consistent with
their EELS data. This implies that on average the a-C:H layer has a
thickness of about 0.3 times the radius of the diamond core (as
schematically shown in Fig.~\ref{section}). For a 0.6~nm radius
grain, such as the one shown in Fig.~\ref{small}, this is equivalent
to about one monolayer of a-C:H or the complete reconstruction to the
sp$^2$ configuration of all the surface C--C bonds in this
particle. For the more typical meteoritic diamond dimensions (radius
$\sim 1.4$~nm, e.g., Fig.~\ref{big}) this would correspond to 2--3
monolayers of a-C:H, or equivalently the outer 2--3 atomic layers
existing in a reconstructed sp$^2$ configuration (e.g.,
Fig.~\ref{section}).

The model presented here rests principally upon the hypothesis that
the surfaces of interstellar nanodiamonds will reconstruct to an
sp$^2$ hydrogenated configuration. The reconstruction process seems in
little doubt (e.g., Blake et al. 1988; Bernatowicz et al. 1990;
Mitsuda et al. 1991; Chin et al. 1992; Dorschner et al. 1996; Hill
1998), however, the role of hydrogen on these nanodiamond surfaces is
not yet clear. On macroscopic diamond crystal faces this surface
reconstruction can be transformed back to a sp$^3$ dominated surface
by hydrogen atom adsorption (e.g., Chin et al. 1992). Whether or not
this same process can occur on a nanodiamond particle in the
interstellar medium is not at all clear. Although, the presence of
meteoritic presolar diamonds with a considerable sp$^2$ surface
component (Bernatowicz et al. 1990) would seem to indicate strongly
that the surface reconstruction is preserved in nanometre-sized
diamonds.

\section{The infrared properties of nanodiamonds}

With the above assumption that the outer few layers of the presolar
nanodiamonds are reconstructed to a sp$^2$ state from their usual
sp$^3$ hybridisation state, we can make some important deductions on
the infrared properties of nanodiamonds. Here we assume that sp$^3$
surface atoms reconstruct to sp$^2$ carbon atoms and that all dangling
carbon bonds are hydrogenated. We now consider the contributions that
nanodiamonds can make to the observed infrared emission from
interstellar dust.

One somewhat intriguing aspect of the observed and unidenti\-fied
inter\-stellar and circum\-stellar infra\-red emission bands in the
3--15~$\mu$m wavelength region is the approximate constancy of the
band positions, widths and the band\--to\--band intensity ratios in
the most common types of infrared emission band spectra (e.g., Geballe
1997, Boulanger 1999). This seems to imply that the carriers of these
bands show little variation in their composition from one region to
another, and over five orders of magnitude in the radiation field
intensity (Boulanger 1999). Aromatic hydrocarbon materials (Duley \&
Williams 1983) and polycyclic aromatic carbons, PAHs, (L\'eger \&
Puget 1984) were long ago proposed as the carriers of the unidentified
infrared emission bands. It has always been somewhat difficult to
explain the apparent constancy and the band widths and positions with
the PAH hypothesis because of the widely varying spectral
characteristics exhibited by these large planar molecular species.
However, recent attempts at matching the emission band spectrum of
several sources (Allamandola et al. 1999) have been more successful. A
coal model for the origin of the emission bands (Papoular et al. 1989)
was also proposed and has provided a good match to the observed
emission band positions and profiles. However, this model too is not
without its problems regarding, in particular, the emission mechanism
(e.g., Puget et al. 1995).

One solution to the conundrum of the uniformity of the emission bands
would be to put the PAH-type of emitting species on the surface of a
three-dimensional particle, which would then greatly reduce variations
in the carrier structures and hence the variability of their emission
spectra. Small grains of non-carbonaceous materials, e.g., silicates
or metal oxides, will not provide the correct lattice structure for
the preservation of the planar sp$^2$ carbon atom surface-coating
structures. Thus, nanodiamond surfaces appear to be the only viable
species with the ability to preserve planar sp$^2$ carbon atom surface
structures on a resilient particle.

\subsection{Bulk features}

Quaternary (4$^\circ$) sp$^3$ carbon atoms in the bulk of a perfect
diamond lattice probably play little role in the observable properties
of nanodiamonds; other than contributing to the overall particle size
and absorption at ultraviolet wavelengths. However, defects within the
bulk due to vacancies or substituted nitrogen atoms could be
responsible for emission features in the $7-30~\mu$m region. Defects
break the lattice symmetry and allow one phonon modes to become active
in this wavelength region. In the perfect diamond lattice only two and
three phonon modes are allowed (e.g., Hill et al. 1998; Hill
1998). Two and three phonon modes in the $3.5-5.5~\mu$m region are
suppressed in nanodiamonds and will be very weak with respect to
surfaces modes and CC and CH features (Hill 1998). Infrared-active
modes in the $6-15~\mu$m region in defective diamonds (Morelli et al.
1993; Hill et al. 1998) may be equivalent to the $6-15~\mu$m infrared
modes expected from nanodiamond surface structures. However, these
will be surface and not bulk features.  As an example of the
importance of bulk modes in interstellar nanodiamond particles Hill et
al. (1998) have shown that a band at $\sim~21~\mu$m in defective
diamonds may be related to the observed $\sim~21~\mu$m emission band
observed in some carbon-rich PPNe.

Defects, such as a vacancies and nitrogen atoms, also exist within the
bulk structure of nanodiamonds. Hill et al. (1997) pointed out that
the anomalous nitrogen component in the meteoritic diamonds is only
liberated at high temperatures in combustion experiments. This
nitrogen must therefore be embedded within the crystallite structure.

In the meteoritic nanodiamonds the nitrogen atom concentration is at
most 13,000~ppm (Russell et al. 1996). As Hill et al. (1997) showed
this is equivalent to only one nitrogen atom in a 1~nm radius
nanodiamond and of the order of $1-10$ nitrogen atoms in a 2~nm
particle. Thus, there are few nitrogen atoms per nanodiamond but their
signatures may still be detectable. They will be very weak in
absorption and probably only detectable in emission. Interestingly,
the nitrogen impurities in the diamond lattice show infrared
absorption bands in the $7-10~\mu$m region. In particular, bands at
8.8, 7.8 and 8.6~$\mu$m due, respectively, to one, two, and four
nitrogen atom substitutions for carbon in the diamond lattice are seen
(e.g., Davies 1977). Braatz et al. (1998) argue, based on their
presolar diamond spectra, that the 8.6 and 8.8~$\mu$m bands are
probably the most likely bands in presolar diamonds.

\subsection{Surface features}

The infrared spectroscopy of presolar nanodiamonds has shown that
surface species dominate the measured spectra, as Hill et al. (1998)
demonstrated. The strongest nano\-diamonds modes will therefore be
those associated with surface groups and structures. Following the
conclusion of Bernatowicz et al. (1990), that the surface layer of the
presolar diamonds appears to be similar to an a-C:H material, we adopt
a-C:H as a model for the nanodiamond surface structure. We then take
the infrared signatures of a-C:H materials (e.g., Robertson 1986) as
an indicator of the infrared properties of reconstructed nanodiamond
surfaces: see next sub-section and the discussions in \S~4. It is
clear that aromatic or olefinic CH and CC modes should dominate and
could be observable as a set of infrared emission bands from particles
heated by the the interstellar radiation field provided that they can
reach an appropriate emission temperature (see \S~6).

Solo and duo CH modes will dominate because many of the surface CH
groups will be isolated. Trio CH groups may also be present at the
particle edges but will be much less abundant than the solo and duo CH
groups. As noted in \S~4 and by Bernatowicz et al. (1990) the surface
composition of nanodiamonds is relatively independent of the particle
size. Therefore the infrared emission bands and the band\--to\--band
intensity ratios should be relatively insensitive to the particle
size.

\subsection{The infrared modes of nanodiamonds}

The reconstructed surfaces of nanodiamonds may be very closely related
to the diamond-like a-C:H materials (Bernatowicz et al. 1990) that
have been extensively studied in the laboratory (e.g., Angus et
al. 1989). Thus, based on the interpretation of the infrared features
seen in laboratory a-C:H materials (e.g., Robertson 1986) we present
in Table~2 some of the expected infrared active modes for
surface-reconstructed nanodiamonds. The reconstruction or
aromatisation of a nanodiamond surface and the formation of sp$^2$
carbon atom CC and CH bonds requires a supression of sp$^3$ surface
carbons. If this process is incomplete the residual sp$^3$ CH bonds
will give rise to aliphatic 3.4 and 3.5~$\mu$m CH features. Hill
(1998) has shown that the $\sim 3.42$ and $\sim 3.53~\mu$m bands of
the diamond precursor molecule adamantane, C$_{10}$H$_{16}$, may be
consistent with the observed emission bands at these wavelengths in
some PPNe, and Guillois et al. (1999) have shown them to be consistent
with $\sim 50$~nm radius diamond particles..

\begin{table}
\caption{The expected infrared modes of surface-reconstructed
 nanodiamond a-C:H surfaces (Bernatowicz et al. 1990) based on the
 laboratory data for a-C:H (Robertson 1986), with the exception of the
 8.6~$\mu$m band assignment which was taken from Allamandola
 (1989). Relative band strengths have not been assigned.}
\begin{tabular}{|c|c|l|}
\hline
\multicolumn{2}{|c}{Band Position} & \multicolumn{1}{|c|}{Origin of the Band}\\
\multicolumn{1}{|c}{$\mu$m}        & \multicolumn{1}{c}{cm$^{-1}$}           & 
\multicolumn{1}{|c|}{ }                                                      \\
\hline 
\hline
\multicolumn{3}{|l|}{C---H stretching modes}                               \\
\hline 
\hline
  3.28       &  3045     &  sp$^2$ aromatic C--H                           \\
\hline
  3.33       &  3000     &  sp$^2$ olefinic C--H                           \\
\hline
  3.42       &  2920     &  sp$^3$ CH \& CH$_2$                            \\
\hline
  3.51       &  2850     &  sp$^3$ CH$_2$                                  \\
\hline 
\hline
\multicolumn{3}{|l|}{C---H bending modes}                                  \\
\hline 
\hline
 8.6         &  1150     &  sp$^2$ aromatic, in-plane                      \\
\hline
 11.36       &  880      &  sp$^2$ solo aromatic, out-of-plane             \\
\hline
 12.2--12.8  & 780--820  &  sp$^2$ duo/trio aromatic, out-of-plane         \\
\hline 
\hline
\multicolumn{3}{|l|}{C---C modes}                                          \\
\hline 
\hline
 6.17        &  1620     &  sp$^2$ olefinic C--C stretch                   \\
\hline
 6.37        &  1570     &  sp$^2$ aromatic C--C stretch                   \\
\hline
 6.63        &  1509     &  ?                                              \\
\hline
 6.99        &  1430     &  sp$^2$ aromatic C--C stretch                   \\
\hline
 7.32        &  1367     &  disorder mode                                  \\
\hline
 7.69        &  1300     &  sp$^3$ C--C stretch                            \\
\hline

\end{tabular}
\end{table}

The tabulated feature assignments given by Robertson (1986) do not
include a band at $\sim 8.6~\mu$m. However, the infrared absorption
spectra presented do show some structure on the long wavelength side
of the band centred at $\sim 8~\mu$m which could be related to a
feature at $\sim 8.6~\mu$m seen in the interstellar medium (e.g.,
Allamandola 1989). The infrared bands seen in laboratory a-C:Hs
(Table~2) thus bear a remarkable wavelength coincidence to the strong
infrared emission bands observed at 3.3, 6.2, 7.7, 8.6, 11.3 and
12.7~$\mu$m and also to the weaker bands at 3.4 and 3.51~$\mu$m (e.g.,
Allamandola 1989). Kapitonov \& Kon'kov (1998) noted this similarity
in the a-C:H bands and proposed these materials as carriers of the
unidentified mid-infrared emission bands. Similarly, Koike et
al. (1995) suggested that diamond-like materials could be one of the
carriers of the minor unidentified infrared bands.

The most commonly observed emission spectra show dominant bands at 3.3,
6.2, 7.7 and 11.3~$\mu$m, weaker bands at 3.4, 5.25 and 8.6~$\mu$m,
and broad plateaux in the $3.3-3.6$ and $11.3-12.6~\mu$m regions
(Geballe 1997). In addition a weak band at 11.0~$\mu$m also seems to
be a common feature of these spectra (e.g., Verstraete et
al. 1996). Without detailed modelling of the of the surface bonds on
nanodiamonds it is clearly premature to attempt to explain all of the
commonly observed emission bands in terms of the emission from these
grains. We do, however, note one interesting fact that may be of
relevance, concerning the position of the solo CH band in the linear
PAHs anthracene, tetracene and pentacene. As the laboratory infrared
spectral catalogues (e.g., Aldrich and Sadtler) show, for this series
of PAHs, the solo CH band occurs at 11.3, 11.1 and 11.0~$\mu$m for the
three, four and five adjacent rings, respectively. For the non-linear
three to five ring PAHs the catalogues show that the solo CH band
always occurs in the $11.2-11.5~\mu$m region. Thus, it appears that as
the number of unperturbed adjacent solo CH sites increases the
11~$\mu$m band shifts to shorter wavelengths. Such a linear
arrangement of solo CH groups is analogous to the structures expected
to form on nanodiamond surfaces (see \S~4 and
Fig.~\ref{section}). Conversely, in the non-linear PAHs the solo CH
sites are usually not adjacent, and are often perturbed by steric
effects due to the location of adjacent CH groups at concavities at
the edge of the PAHs. These sites would be analogous to the CH groups
at, and near, steps and edges on a nanodiamond surface. Thus, as a
natural consequence of the surface structure of nanodiamond particles,
the presence of a weak band at 11.0~$\mu$m and the origin of the more
prominent 11.3~$\mu$m band can be explained.

\section{The thermal properties of nanodiamonds}

The meteoritic nanodiamonds provide a model for the properties of the
stochastically-heated interstellar grains with sizes intermediate
between PAH molecules and large grains, i.e., the so-called Very Small
Grains (VSGs, D\'esert et al.  1990). The presolar nanodiamonds are,
in effect, core/mantle particles with cores that absorb strongly at
ultraviolet wavelengths, and mantles that emit at infrared
wavelengths. Thus, because the heat capacity per unit volume of
diamond is greater than that of graphite and a-C:H, and because
diamond is a weak infrared emitter, the diamond core will act as a
thermal reservoir that pumps infrared emission from the surface
material.

We have estimated the thermal properties of the meteoritic diamonds in
order to calculate the maximum temperatures of these particles upon
absorption of a single photon from the interstellar radiation field.
In this calculation we assumed the nanodiamond structure of
Bernatowicz et al. (1990), i.e., 46~\% by volume of fcc diamond and
54~\% by volume of a-C:H in the form of a coating. For the diamond
heat capacity we used the data from Lide (1992) and adopted a Debye
temperature, $\theta$, of 2034.4~K to extrapolate these data to lower
temperatures using $C(T)_d \propto (T/\theta)^3$ smoothly fitted to
the Lide (1992) data at 300~K. As we do not know the heat capacity of
the a-C:H coating material we have adopted the heat capacity for
graphite from Lide (1992) and extrapolated this to lower temperatures
using a Debye temperature of 1819.7~K, in the same way as for the
diamond core. However, in this case we make allowance for the lower
density of a-C:H, with respect to graphite, by multiplying the derived
heat capacity by the ratio of the specific densities,
($\rho_a/\rho_g$), where $\rho_a$ and $\rho_g$ are the specific mass
densities of a-C:H and graphite, respectively. We also make allowance
for the fact that the particle surface is hydrogenated and that these
hydrogen atoms can contribute to the heat capacity. For the CH bond
heat capacity, $C_H(T)$, we use the values given by Dwek et
al. (1997). The specific heat capacity of a nanodiamond particle, as a
function of its radius $a$, can therefore be estimated using the
following expression,

\[
C(T) = [\ 0.46\ N(a)_d\ C_C(T)_d\ + 
\]
\[
\ \ \ \ \ \ \ \ \ \ \ \ \ 0.54\ N(a)_a\ (\rho_a/\rho_g)\ C_C(T)_g\ + 
\]
\begin{equation}
\ \ \ \ \ \ \ \ \ \ \ \ \ N_{sH}(a)_a\ C_H(T)\ ]\ /\ \frac{4}{3}\pi a^3\ ,
\end{equation}
where, $N(a)_d$ and $N(a)_a$ are the number of atoms in the dia\-mond
and a-C:H phases and $C_C(T)_d$ and $C_C(T)_g$ are the heat capacities
per carbon atom for diamond and graphite, respectively. $N_{sH}(a)_a$
is the number of surface hydrogen atoms on the a-C:H mantle (we have
assumed 50~\% surface coverage in our calculations).

\begin{figure}
 \resizebox{\hsize}{!}{\includegraphics{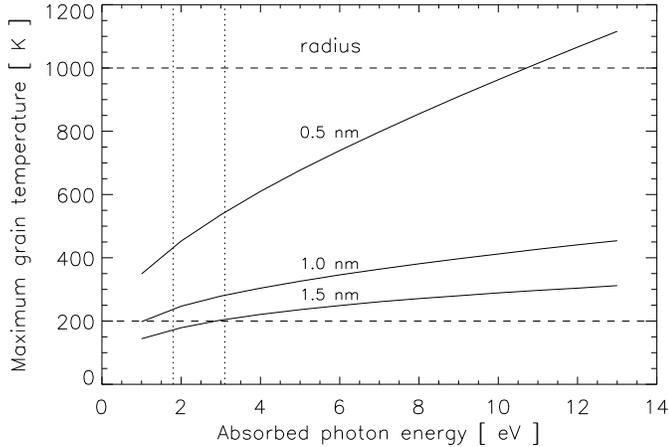}}
 \caption{Nanodiamond maximum temperatures as a function of particle
 size and absorbed photon energy. The horizontal dashed lines mark the
 approximate range of dust temperatures required to produce the
 infrared emission bands ($\sim 200$ to 1000~K). The vertical dotted
 lines mark the visible photon energy range.}
 \label{Tmax}
\end{figure}

In Fig.~\ref{Tmax} we show the maximum nanodiamond temperature,
$T_{max}$, as a function of the particle size and the absorbed photon
energy, $E$, calculated by solution of the equation,
\begin{equation}
E = \frac{4}{3} \pi a^3 \int_{T_0}^{T_{max}} C(T) dT\ ,
\end{equation}
where, $T_0$ is the grain temperature prior to photon absorption which
we have taken to be zero. For grains to emit efficiently at
wavelengths in the $3-15~\mu$m region they must have temperatures in
the range $200-1000$~K. From Fig.~\ref{Tmax} we can see that
emission in the $3-15~\mu$m region from nanodiamond surface sp$^2$ CC
and CH groups requires grain radii in the $0.5-1.5$~nm range, i.e.,
sizes entirely consistent with the presolar meteoritic diamonds.

We therefore find that small interstellar nanodiamonds, with radii of
the order of 0.5~nm, will likely have maximum temperatures of the
order of 1000~K for ultraviolet photon absorption.  Such high
temperatures are likely to eject loosely bound surface species and
thus continually scrub the grain surfaces clean of weakly adsorbed
atoms or molecular groups. This process may therefore provide the
means to preserve the surface structure and composition of the
interstellar nanodiamonds.

\begin{figure}
 \resizebox{\hsize}{!}{\includegraphics{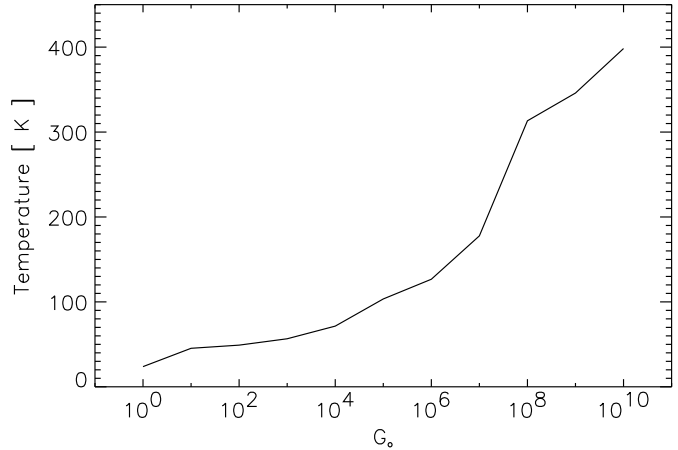}}
 \caption{Nanodiamond equilibrium temperatures as a function of the
 intensity of the interstellar radiation field, in units of the
 standard local radiation field in the Solar neighbourhood, $G_0$. In
 this regime, particle size $\ll$ wavelength, the nanodiamond
 temperatures are independent of the radius.}
 \label{Teq}
\end{figure}

In Fig.~\ref{Teq} we show the equilibrium temperatures of
nanodiamonds as a function of the radiation field in units of $G_0$,
where $G_0$ is the radiation field in the Solar neighbourhood. The
data in Fig.~\ref{Teq} were calculated using $Q_{abs}$ values in
the long wavelength limit derived for core/mantle grains using the
a-C:H refractive index data (sample BE of Rouleau \& Martin 1991)
and the diamond refractive index data from Roberts \& Walker
(1967). Note that in Fig.~\ref{Teq} the nanodiamond equilibrium
temperatures are much less than the maximum temperatures shown in
Fig.~\ref{Tmax}. Stochastic heating will therefore be the dominant
heating process for nanodiamonds in almost all radiation field
environments. However, if the calculated equilibrium temperatures, for
$G_0 < 10^4$ (i.e., $\sim 25-70$~K), are indicative of the mean
nanodiamond temperatures, then this implies continuum emission from
these grains in the $\sim 40-120~\mu$m region.

Infrared continuum emission from nanodiamonds will probably be
sensitive to the thermal connectivity between the bulk atoms and the
surface structures. In low excitation regions emission from the
surface CC and CH modes will dominate in the mid-infrared because the
grain thermal energy will be channelled through these modes. In high
excitation regions the bulk may also contribute to the emission at the
same time as continuing to pump the surface emission modes. Thus, at
higher grain temperatures the bulk infrared bands in the diamond one
phonon region, i.e., in the $10-15$ and $\sim 21~\mu$m bands, may also
be seen in emission (Hill et al. 1998). The diamond two and three
phonon nanodiamond modes in the $3.5-5.5~\mu$m region may also appear
as two weak emission bands centred at $\sim 4~\mu$m and $\sim 5~\mu$m
(Hill 1998).  Interstellar nanodiamonds heated to temperatures of the
order of 1000~K (Fig.~\ref{Tmax}) may also be responsible for
continuum emission at wavelengths of the order of a few microns.

The grain temperature arguments presented here are
consistent with emission features in the mid-infrared from surface CC
and CH groups on nanodiamond surfaces. However, because of
uncertainties in the details of the calculation of the specific heat
capacities of nanometre-sized amorphous particles these conclusions
need to be verified experimentally. Clearly, in order for nanodiamonds
to be responsible for the mid-infrared emission bands their sizes must
extend down to sizes of 0.5~nm. In the log-normal size distribution of
the presolar diamonds in the Murchi\-son meteorite particles with
radii small\-er than 0.5~nm seem to be present (Daulton et
al. 1996). The presolar diamond size distribution therefore seems to
be entirely consistent with the above requirement for small
nanodiamonds (radii $= 0.5 - 1.5$~nm) to explain the mid-infrared
emission bands.

\section{Nanodiamonds in the interstellar medium}

Nanodiamonds are very stable and resilient species: they can survive
the passage of supernova-generated shocks.  Only in very fast shocks
(velocities in excess of 200~km~s$^{-1}$) will they be eroded by
thermal sputtering in the more than million degree post-shock gas
(Jones et al. 1996,1997). The need for a very refractory carbonaceous
interstellar dust material is indicated by observations which show an
almost constant abundance of gas phase carbon over a wide range of
molecular hydrogen abundance (Cardelli et al. 1996; Sofia et
al. 1997).  Thus, indicating little exchange of carbon between the
gaseous and solid phases in the interstellar medium. This would be
consistent with a large fraction of the solid carbon being in a very
refractory phase such as diamond.

\subsection{Visible and ultraviolet absorption and emission}

Presolar nanodiamonds in the laboratory are a pale grey or brown
colour and therefore absorb visible radiation. Then by implication,
infrared emission bands arising on nanodiamond surfaces will be
excited by visible photons. Such absorption is required to explain the
emission bands observed in the reflection nebula vdB 133 (Uchida et
al. 1998), the galaxy M31 (Cesarsky et al. 1998; Pagani et
al. 1999), and elliptical galaxies (Madden et al. 1999), where there
is little ultraviolet radiation.

Duley (1988) noted a remarkable similarity in the sharp emission
features observed in the luminescence spectra of terrestrial diamonds
and those seen in the Red Rectangle. This observation therefore seems
to lend further support to the presence of diamonds in the
interstellar medium.

Sandford (1996) suggested that the meteoritic diamonds could be the
carrier of the 217.5~nm extinction feature. His argument was that
complete surface reconstruction to sp$^2$ carbon will de-localise the
$\pi$ electrons (e.g., Fig.~\ref{section}) and lead to a surface
plasmon mode analogous to that of graphite.  Diamond absorbs strongly
at ultra\-violet wavelengths and therefore nanodiamonds should
contribute to the ultraviolet extinction (Lew\-is et al.  1989).

\subsection{Observational constraints}

In the model proposed here one type of particle, namely
surface-reconstructed nanodiamonds, is invoked as the carrier of three
different observed interstellar dust properties; the infrared emission
bands (a property of the sp$^2$ carbon surface CC and CH groups), the
ultraviolet extinction bump (a property of the sp$^2$ surface
connectivity) and a contribution to the ultraviolet extinction (a
property of the overall grain radius and optical properties). The
first two properties are related to different surface effects and the
third is due to the particle size and absorptivity. These three
observable properties of nanodiamond may therefore not be well
correlated, even though they arise from the same particles. A lack of
correlation between the 217.5~nm extinction bump and the ultraviolet
extinction is clearly required by the observational constraints (e.g.,
Mathis 1990). However, a correlation between the area of the
extinction bump and the strength of the infrared emission bands has
been observed (Boulanger et al. 1994) which would be consistent with
both features arising in the same type of particles.

Recent tight observational constraints on the abundance of carbon in
the interstellar medium now place severe limitations on interstellar
dust models (e.g., Snow \& Witt 1995, 1996). Thus, any dust model that
uses one material, in the same dust particles, to explain several of
the observed properties of interstellar dust will more easily be able
to meet the strict abundance constraints. Interstellar nanodiamonds as
an abundant and important grain component, would therefore certainly
help to resolve this problem.

\section{Summary}

We propose that nanodiamonds are an important component of
interstellar dust, and that far from being difficult to observe we
have infact long been observing them. We suggest, principally based on
laboratory work on the morphological properties of presolar
nanodiamonds, that they may be a
major contributor to the unidentified interstellar mid-infrared
emission bands in the $3-15~\mu$m region. The emission from
nanodiamonds in interstellar and circumstellar media will be driven by
the stochastic absorption of visible and ultraviolet photons leading
to peak temperatures in the range $\sim 100-1000$~K.

It seems apparent that this model inherently encompasses aspects of
both the PAH (e.g., the specific polyatomic emitting species) and coal
model hypotheses (e.g., the three-dimensional nature of the carrier),
and may thus provide a bridge between these two research fields.

The nanodiamond hypothesis presented here naturally explains the
positions of the emission bands, the dominance of the 11.3~$\mu$m solo
CH feature, the presence of a band at $11.0~\mu$m, and the relative
invariance of the emission spectra and the band\--to\--band intensity
ratios. These are explained by a nanodiamond surface origin for the
bands and by the fact that the structural properties of nanodiamond
surfaces are relatively insensitive to the particle size.

In addition to making a major contribution to the mid-infra\-red
emis\-sion bands, inter\-stellar nanodiamonds may also be the source
of the 217.5~nm extinction bump and may make an important contribution
to the ultraviolet extinction.

One interesting and useful outcome of this hypothesis is that it can
lead to an easing of the tight constraints placed on interstellar dust
models by putting three dust properties into one particle, namely, the
infrared emission bands, the 217.5~nm extinction bump and a
contribution to the ultraviolet extinction. The fact that these three
dust emission and extinction properties do not seem to be well
correlated is consistent with the hypothesis. This arises because the
three properties arise from different aspects of the particle, namely
surface molecular groups, the continuity of the sp$^2$ surface and the
particle size, respectively.

The infrared band carriers and the VSGs in this hypothesis are one and
the same species. The characteristics of each type of particle,
derived from the interpretation of observational data, are just
different aspects of the emission from one type of particle. The
smaller particles, stochastically heated to high temperatures, produce
the emission bands, and the cooler larger particles produce underlying
plateaux/continua. In high excitation regions the infrared emission
band-to-continuum ratio should decrease as the grain core increasingly
contributes to the continuum emission. Other bands in the $\approx 4$,
5, $10-15$ and 21~$\mu$m regions may also be seen in emission in high
excitation regions.  It is also likely that the smallest nanodiamond
grains, stochastically heated to temperatures of the order of 1000~K,
could give rise to an infrared continuum in the $2-3~\mu$m wavelength
region.

Clearly, in order to fully exploit and develop the proposed
interstellar nanodiamond hypothesis new experi\-mental and
theoret\-ical developments are required. However, both of these fields
will be difficult to pursue in the nanometre-size particle regime. The
experiments will be hampered by the very active surface chemistry of
nanodiamonds, and by the need to isolate and study such small
particles. Theo\-retical developments will rely on modelling particles
with of the order of hundreds to thousands of atoms using some form of
a quasi-quantum mechanical approach, rather than trying to extrapolate
bulk properties to nano\-metre size scales which is clearly
inappropriate.

\begin{acknowledgements}
We thank H. Hill, F. Boulanger, L. Verstraete and S. Madden for
fruitful discussions on the nature of nanodiamonds and interstellar
grains.
\end{acknowledgements}



\end{document}